\documentclass[namedreferences]{solarphysics}

\usepackage[optionalrh]{spr-sola-addons} % For Solar Physics 
\usepackage{graphicx}        % For eps figures, newer & more powerfull
\usepackage{color}           % For color text: \color command
\usepackage{breakurl}        % For breaking URLs easily trough lines
\usepackage{bm}

\newcommand{\alphabold}{\mbox{\boldmath$\alpha$}}

\newcommand{\vect}[1]{\bm{#1}}

\newcommand{\aap}{    {\it Astron. Astrophys.}}

\newcommand{\apjl}{   {\it Astrophys. J. Lett.}}

\chardef\us=`\_

\begin{document}

\begin{article}
\begin{opening}

\title{Accelerating Multiframe Blind Deconvolution via Deep Learning}

\author[addressref={aff1,aff2},email={andres.asensio@iac.es}]{\inits{A.}\fnm{Andr\'es}~\lnm{Asensio Ramos}}
\author[addressref={aff1,aff2},email={sara.esteban@iac.es}]{\inits{S.}\fnm{Sara}~\lnm{Esteban Pozuelo}}
\author[addressref={aff1,aff2},email={ckuckein@iac.es}]{\inits{C.}\fnm{Christoph}~\lnm{Kuckein}}
%\author[addressref=aff3,email={nolspert@gmail.com}]{\inits{N.O.}\fnm{Nigul}~\lnm{Olspert}}

%\author{\inits{}\fnm{}~\lnm{}\orcid{}}
%   NOTE:  Just one corresponding author [corref]
%   \institute{$^{1}$ First affiliation
%                     email: \url{e.mail-a} email: \url{e.mail-b}\\ 
%              $^{2}$ Second affiliation
%                     email: \url{e.mail-c} \\
%             \textit{}

\address[id=aff1]{Instituto de Astrofisica de Canarias, 38205, La Laguna, Tenerife, Spain}
\address[id=aff2]{Departamento de Astrofisica, Universidad de La Laguna, E-38205 La Laguna, Tenerife, Spain}
\address[id=aff3]{Max-Planck-Institut fur Sonnensystemforschung, Justus-von-Liebig-Weg 3, 37077 Gottingen, Germany}

\runningauthor{Asensio Ramos et al.}
\runningtitle{Accelerating multiframe blind deconvolution via deep learning}

\begin{abstract}
Ground-based solar image restoration is a computationally expensive procedure
that involves nonlinear optimization techniques. The presence of
atmospheric turbulence produces perturbations in individual
images that make it necessary to apply blind 
deconvolution techniques. These techniques rely on the observation of many short exposure frames
that are used to simultaneously infer the instantaneous state of the atmosphere and the unperturbed
object. We have recently explored the use of machine learning to
accelerate this process, with promising results. We build upon this
previous work to propose several interesting improvements that lead to
better models. As well, we propose a new method to accelerate the restoration based on algorithm unrolling. In
this method, the image restoration
problem is solved with a gradient descent method that
is unrolled and accelerated aided by a few small 
neural networks. The role of the neural networks 
is to correct the estimation of the solution at
each iterative step. The model is trained to perform the
optimization in a small fixed number of steps with a curated dataset. Our findings demonstrate 
that both methods significantly reduce the restoration time compared to 
the standard optimization procedure. Furthermore, we showcase that these models can be trained 
in an unsupervised manner using observed images from three different instruments.
Remarkably, they also exhibit robust generalization capabilities when applied to new datasets. 
To foster 
further research and collaboration, we openly provide the trained models, along with the 
corresponding training and evaluation code, as well as the training dataset, to the scientific community.
\end{abstract}
\end{opening}
%-------------------------------------------------

\section{Introduction}
\label{sec:Introduction} 

Ground-based solar observations are affected by aberrations caused by the 
turbulent nature of Earth's atmosphere. Although moving the telescope to space is
feasible, it is expensive. That is the reason why correction methods have been
developed to improve ground-based observations. Real-time adaptive optics (AO)
measure the distortion of the wavefront using wavefront sensors in real-time and try to compensate for it
using deformable mirrors. Once measured, a-posteriori software methods push the
quality further.

Although AO systems can remove a substantial amount of aberrations in real-time, they still
suffer from limitations. The corrections are not perfect because of the lag between
the measurement of the wavefront and the correction. This leads to some perturbations remaining in the 
science images, which cannot reach the diffraction limit of the telescope. Additionally, 
single-conjugate AO systems, which are available in the majority of solar telescopes, 
only correct the turbulence in the pupil of the telescope, whose influence is the same
for the whole field-of-view (FoV). The presence of turbulence at higher layers
in the Earth's atmosphere produces differential aberrations in different regions of
the FoV. 

To achieve the diffraction limit of the telescope, it is crucial to remove 
residual aberrations, which can only be practically achieved with a-posteriori 
correction methods. Since neither the instantaneous state of the atmosphere 
nor the unaberrated object are known, all a-posteriori methods belong to the
class of blind deconvolution algorithms. In these algorithms, prior information
is used to simultaneously solve for the object and the atmospheric aberrations,
a problem that is known to be ill-defined. The application of these methods requires specific observing
strategies. The main requisite is to observe a sufficiently large burst of images
of low quality that can later be combined to produce an image of improved quality.
One requires the exposure time of these individual frames to be faster than the coherence
time of the atmosphere in order to freeze the atmosphere. For a typical observing
day, the integration time needs to be as small as a few milliseconds. Although
of excellent performance, unfortunately, these methods are computationally demanding. They
require the maximization of a likelihood (or posterior if priors are used) 
with respect to the instantaneous point spread functions (PSF) and the object.

Arguably the simplest a-posteriori correction method is that of lucky imaging or frame
selection. This requires the definition of a metric measuring the quality of each
individual image and only selecting the best one that maximizes the metric. Although
effective in some cases \citep[e.g.,][]{fastcam08}, they make a very limited
use of the available information because they do not combine low-quality
frames for the sake of producing an improved image. More elaborate techniques, like
speckle methods \citep{labeyrie70,vonderluhe93}, do make use of all frames
and try to estimate a high-quality image exploiting the statistics
of the atmospheric turbulence. Arguably, a step forward happened when 
\cite{paxman92} proposed a blind deconvolution method based on three main pillars. The
first one is a consistent model for the PSFs, which is of special interest for telescopic
observations. This model is obtained by a linear expansion of the wavefront
in the pupil in a suitable basis (Zernike polynomials and Karhunen-Lo\`eve 
modes\footnote{Karhunen-Lo\`eve (KL) modes are obtained after a suitable
rotation of the Zernike basis by diagonalizing the covariance matrix
under the assumption of Kolmogorov turbulence \citep{noll76}.} are
the two most used basis). The PSF is obtained as the autocorrelation of
the pupil function (see section \ref{sec:mfbd} for more details), which 
naturally leads to positive defin
ite PSFs in which
diffraction is automatically taken into account. The second pillar is
a proper treatment of the noise through a Bayesian approach. The function
to be optimized, the likelihood, depends on the assumed noise statistics. Finally, 
the third pillar is the use
of phase-diversity techniques \citep{gonsalves79} to better inform the optimization.
The idea behind phase-diversity is to simultaneously acquire an image with
a known aberration, which is used to constrain the optimization.
Based on this approach, \cite{lofdahl_scharmer94} and \cite{lofdahl98} applied
the phase-diversity (image correction with only two simultaneous images, one at focus and the 
other one with a known defocus) approach to solar observations. Later, \cite{lofdahl02} 
proposed to use many short exposure frames (multi-frame approach) to solve the
blind deconvolution problem. Each frame is affected by different aberrations
although they come from the same object. Both approaches were very successful, and
were later combined and expanded by \cite{vannoort05}. These
authors considered, apart from having many frames with phase diversity, the
presence of many objects (multiobject), typically made of observations at different wavelengths. 
These objects are obtained simultaneously and assumed to be affected by the
same aberrations. They developed the multiframe multiobject blind deconvolution 
code (MOMFBD), which is now routinely used in solar observations.

Given the high computational requirements of multi-frame (multi-object) blind deconvolution
classical methods, \cite{AsensioRamos2018} investigated the use of learning-based methods that 
leverage neural networks to correct atmospheric influences quickly. They proposed 
two large convolutional neural networks (CNNs) to correct a burst of images and 
produce a frame of higher quality. The CNNs were trained supervisedly with 
observations of the Swedish 1-m Solar Telescope \citep[SST;][]{2003SPIE.4853..341S}, which were previously 
corrected with the MOMFBD code. This approach can correct observations as large as 1k$\times$1k at a cadence of 100 
images per second, close to real-time, neglecting the mandatory standard data 
reduction. Although this approach is promising for real-time image correction, 
it has two main disadvantages. The first one is that the training requires a previous run of the
time-consuming MOMFBD code. The second is that only corrected images are obtained, and no information 
about the instantaneous wavefront (or PSF) is inferred, which may prevent correcting other 
simultaneous data obtained at other wavelengths.

Later, \cite{AsensioRamos2020} realized that a neural model can be trained 
simply from corrupted images by making use of the physics of image formation inside the 
model. The neural network is then responsible for predicting the instantaneous 
PSFs, instead of the corrected images. Leveraging the linear theory of image 
formation and using the loss function utilized in the MOMFBD code, which only depends 
on observed corrupted images, the training can be done in a fully unsupervised manner. Apart from 
the advantage of unsupervised training, this approach also allows the PSFs
to be reused for correcting other instruments observing in perfect synchronization in the same 
FoV. With the advent of new telescopes, using several instruments 
to observe the same FoV at the same time will become widespread.

It seems obvious that machine learning-assisted deconvolution is one of the possible avenues
for improving the quality and speed of solar blind deconvolution. This is of special
relevance given the large amount of large-scale observations that we are currently gathering
with current telescopes like the SST at the Observatorio del 
Roque de los Muchachos (Spain), the GREGOR telescope \citep{GREGOR2012,kleint2020} at 
the Observatorio del Teide (Spain), 
the Goode Solar Telescope (GST) on the Big Bear Observatory (USA) or the Daniel K. Inouye
Solar Telescope on the Haleakala Observatory (USA). Future telescopes, like
the European Solar Telescope (EST), will produce an
even larger amount of data that needs to be corrected quickly.

In this paper, we propose two novel ideas. The first one is a set of improvements over the
model proposed by \cite{AsensioRamos2020}, which arguably produces better results. The
second one is an approach for image reconstruction based on algorithm unrolling. This
approach is, in spirit, much closer to the classical MOMFBD optimization approach.

\section{Multi-frame Blind Deconvolution}
\label{sec:mfbd}
The formalism followed in this paper is similar to that of \cite{AsensioRamos2020}, although
with some small but important differences. Let us assume that we 
collect $J$ short exposure frames with a ground-based 
instrument of a stationary object outside Earth's atmosphere. In our case, given our
scientific interests, our object is a small region on the surface of the Sun. The exposure time 
of each individual frame is short enough --of the order of milliseconds-- so that one can assume that the atmosphere 
is ``frozen'', i.e., it does not evolve during the integration time. Additionally, the total duration of 
the burst of exposures has to be shorter than the solar evolution timescales, so that the
object can be safely assumed to be the same for all frames.
From the linear theory of image formation, for a spatially invariant system, the 
image $i_{kj}$ registered on a sensor can be expressed as a convolution of the true object $o_k$ with the
PSF of the optical system and atmosphere, that we denote as $s_{kj}$, plus photon noise $n_{kj}$:
\begin{equation}
i_{kj}(r) = o_k(r) * s_{kj}(r) + n_{kj}(r),
\label{eq:gen_model}
\end{equation}
where $r=(x, y)$ represents the coordinates on the image plane and $j \in \{1 \dots J\}$ 
labels each individual
frame for object $k\in \{1 \dots K\}$. We make the assumption that all $K$ objects are
observed strictly
simultaneously, so that they share the same atmospheric perturbations.
In the standard case of photon noise, the noise can be assumed to be Poissonian. However, in the regime
of high signal-to-noise ratio (S/N), like in the case of our observations, it rapidly converges to 
a Gaussian statistics. 
This assumption greatly simplifies the solution of the problem, as shown below.

The task of MOMFBD is to restore the object $o$ from the measurement of all frames $i_j$. However, the
problem is indeed blind, since both $o$ and all $s_j$ are unknown. Such problem can only
be solved in the Bayesian framework. From a Bayesian perspective, our aim is to compute
the posterior distribution over the object and the PSFs conditioned on the observations, 
$p(\mathbf{o},\mathbf{s}|\mathbf{i})$, which can be computed as:
\begin{equation}
    p(\mathbf{o},\mathbf{s}|\mathbf{i}) \propto p(\mathbf{i}|\mathbf{o},\mathbf{s}) p(\mathbf{o},\mathbf{s}),    
\end{equation}
where $p(\mathbf{i}|o,\mathbf{s})$ is the likelihood function and $p(o,\mathbf{s})$ is the 
prior distribution. The likelihood function takes into account the information encoded in the
data, while the prior encodes all explicit a-priori knowledge about the object and the PSFs. One widespread
solution to the blind deconvolution problem is the one that maximizes the posterior, commonly
known as maximum a-posteriori (MAP) solution:
\begin{equation}
    \arg \max_{\mathbf{o},\mathbf{s}} p(\mathbf{i}|\mathbf{o},\mathbf{s}) p(\mathbf{o},\mathbf{s}). 
    \label{eq:map}
\end{equation}
Prior information can also be incorporated implicitly. In general, one can define a flat prior in the
pixel space and allow the optimization to find the instantaneous PSFs. However, this does not
work well in practice because of the enormous freedom. For this reason, it is
useful to reduce the space of PSFs in which the optimization will search. Following the standard approach
in the solar application of MOMFBD, we regularize the solution by assuming 
that the PSF is obtained via the wavefront entering the pupil of the telescope. We introduce
the generalized pupil function $P_j$, so that the PSF can be obtained as the autocorrelation
of the generalized pupil function:
\begin{eqnarray}
P_{kj}(v)&=&A_k(v)e^{i \varphi_{kj}(v)},\\
s_{kj} &=& |\mathcal{F}^{-1}(P_{kj})|^2,
\label{eq:pupil_func}
\end{eqnarray}
where $A(v)$ is the mask of the telescope aperture (that includes the shadow of the secondary
mirror or any spider if present), $\varphi_{kj}$ is the phase of the wavefront, 
$v$ is the coordinate on the pupil plane and $\mathcal{F}^{-1}$ denotes the 
inverse Fourier transform. Note that $i$ refers to the imaginary unit number.

% In this study we use pairs of images taken simultaneously. One image is taken with the camera
% in focus ($\delta_j(v)=0$) and the other one is taken with a known defocus ($\delta_j(v) \neq 0$). This is known as
% phase diversity (Refs), and is known to strongly regularize the deconvolution
% problem [REF]. For simplicity of notation, instead of using two Equations (\ref{eq_image_formation}) with different
% values of $\delta_j$ for each $j$, it is easier to assume that the index $j$ runs over both atmospheric 
% realizations and diversity channels, so that $j \in \{1 \dots 2J\}$. 

As mentioned above, the wavefronts are parameterized through suitable basis functions. In this study we use the
Karhunen-Lo\`eve basis \cite[e.g.,][]{vannoort05}, which makes the expansion coefficients statistically 
independent under the assumption of Kolmogorov turbulence. This can potentially help
regularize the inversion problem by making the coefficients of the wavefront more
independent. The wavefront can be expressed as \citep{vannoort05}:
\begin{equation}
\varphi_{kj}(v)=\sum_{l=1}^{M}\alpha_{jl}{\rm KL}_l / \lambda_k,
\label{eq:wavefront}
\end{equation}
where $M$ is the number of basis functions and $\alpha_{jl}$ is the coefficient associated with
the basis function $\mathrm{KL}_l$ of the $j$-th atmospheric frame. In all experiments carried
out in this paper, we use $M=44$, which gives excellent results in the recovery of the object.
In all the results shown in this paper, all objects are observed very close in wavelength. This
simplifies the problem and one can drop the wavelength normalization in Eq. (\ref{eq:wavefront}).

We assume flat priors for the KL coefficients and the object $o$. We also assume that
all measured frames $\mathbf{i}$ are statistically independent, so that the likelihood
function factorizes. Taking into account the generative model of Equation (\ref{eq:gen_model}), the solution
of Equation (\ref{eq:map}) can be obtained by optimizing the following quantity (commonly known as loss):
\begin{equation}
    L(o,\alphabold) = \sum_{k,j,r}\gamma_{kj}\left[i_{kj}(r)-o_k(r)* s_{kj}(r, \vect{\alpha}_j)\right]^2,
    \label{eq:loss}
\end{equation}
where $\gamma_j$ is the inverse variance of the photon noise and the summation over $r$ is over all pixels of 
the image. The same log-likelihood can be computed in the Fourier domain, which significantly simplifies
it by removing the convolution:
\begin{equation}
    L(o,\alphabold) = \sum_{k,j,u}\gamma_{kj}\left[I_{kj}(u)-O_k(u)  S_{kj}(u, \vect{\alpha}_j)\right]^2,	
    \label{eq:loss_fourier}
\end{equation}
where, in this case, $u$ represents frequencies in the Fourier domain, $I_{kj} = \mathcal{F}(i_{kj})$ and
$S_{kj} = \mathcal{F}(s_{kj})$, the latter being known as the optical transfer function (OTF).

The simultaneous optimization of the loss functions of Equations (\ref{eq:loss}) or (\ref{eq:loss_fourier}) with respect to the
object and the PSFs is complicated because of its nonconvexity. To this end, we use an alternating
optimization method, already used in this field by \cite{paxman92}. If the wavefront coefficients are known, the 
loss function is linear in the object and can be analytically estimated using, for instance, a Wiener filter:
\begin{equation}	
	\overline{O}_k(u) = \frac{\sum_j \gamma_{kj} I_{kj}(u)S^{*}_{kj}(u, \vect{\alpha}_j)}{\sum_j \gamma_{kj}|S_{kj}(u, \vect{\alpha}_j)|^2+ 
	\frac{S_n}{S_0(u)}},
	\label{eq:wiener}
\end{equation}
where the overline indicates that this is an estimated quantity, while $S_n$ is the noise power 
spectrum and $S_0(u)$ is the estimated power spectrum of the object. Arguably, the simplest
version of the Wiener filter assumes that $S_n/S_0(u)$ is constant and independent
of the frequency. Although this gives good results, it was demonstrated
by \cite{paxman92} and \cite{lofdahl_scharmer94} that improved images can be obtained by applying
a Fourier filter that is a function of the Fourier coordinates $u$:
\begin{equation}
	\overline{O}_k(u) = H_k(u)\frac{\sum_j \gamma_{kj} I_{kj}(u)S^{*}_{kj}(u, \vect{\alpha}_j)}{\sum_j \gamma_{kj}|S_{kj}(u, \vect{\alpha}_j)|^2},
 \label{eq:object}
\end{equation}
being $H$ a filter with the following form:
\begin{equation}
	H_k(u) = 1 - \frac{\sum_j |S_{kj}(u, \vect{\alpha}_j)|^2 }{|\sum_j I_{kj}(u) S_{kj}(u, \vect{\alpha}_j)|^2},
\end{equation}
where all values below 0.2 are set to zero and all values above 1 are set to one. Finally, all
isolated peaks in the filter that are not connected with the peak at $u=0$ are removed after 
applying a median filter with a $3 \times 3$ kernel.

Once the estimated objects are obtained, the loss function of Equation (\ref{eq:loss}) does only
depend on the wavefront coefficients and can be more easily optimized. To this end, one can use
any optimizers for a nonlinear scalar function.

\section{The Models}
\label{sec:models}
We propose here two neural approaches to accelerate the blind deconvolution problem
in extended objects like the Sun. The first one is based on algorithm unrolling. The 
second one is an improvement over the model proposed by \cite{AsensioRamos2020}, in
which the recurrent layers are replaced by convolutional layers and a much
deeper convolutional encoder is used to extract features from the images.
We warn the reader that we do not claim that the architectures described in the following are optimal
in terms of simplicity, efficiency, and accuracy. 

\subsection{Model 1: Algorithm Unrolling}
\label{sec:unrolling}
Algorithm unrolling \citep{fasta10} consists of serializing (or unrolling) any iterative
method for a fixed number of iterations and treating it as a deep model. In principle,
algorithm unrolling can be applied to any iterative algorithm and is often used
with computational purposes in mind, because certain operations can be parallelized
or fused depending on the specific hardware in which it is run. However, it can also
be used to improve the convergence of the algorithm if
neural networks are placed at certain positions in the unrolled algorithm. Since
the majority of optimization algorithms and neural networks are differentiable, 
the resulting model can be trained using backpropagation. This approach
provides a transparent way of mixing together classical iterative
algorithms and neural networks \citep[see][for a recent review]{unrolling21}.

Following the steps described in the previous section, we propose to solve 
Equation (\ref{eq:loss}) repeating the following scheme a fixed number of times $K$:
\begin{itemize}
	\item Assume that the wavefront coefficients are known and estimate the object using 
	Equation (\ref{eq:object}).
	\item Assume that the object is known and carry out a correction of the wavefront coefficients
	using a gradient descent step. This requires the computation of the gradient of the loss function
	with respect to the wavefront coefficients:
	\begin{equation}
		\alphabold_{i+1} = \alphabold_{i} - \eta_i \nabla L(o,\alphabold_i),
		\label{eq:gradient_descent}
	\end{equation}
	where the gradient is taken with respect to $\alphabold$ and $\eta_i$ acts
	as a learning rate. This gradient can be computed
	analytically using the chain rule or using automatic differentiation.
\end{itemize}

\begin{figure*}
	\includegraphics[width=1.0\textwidth]{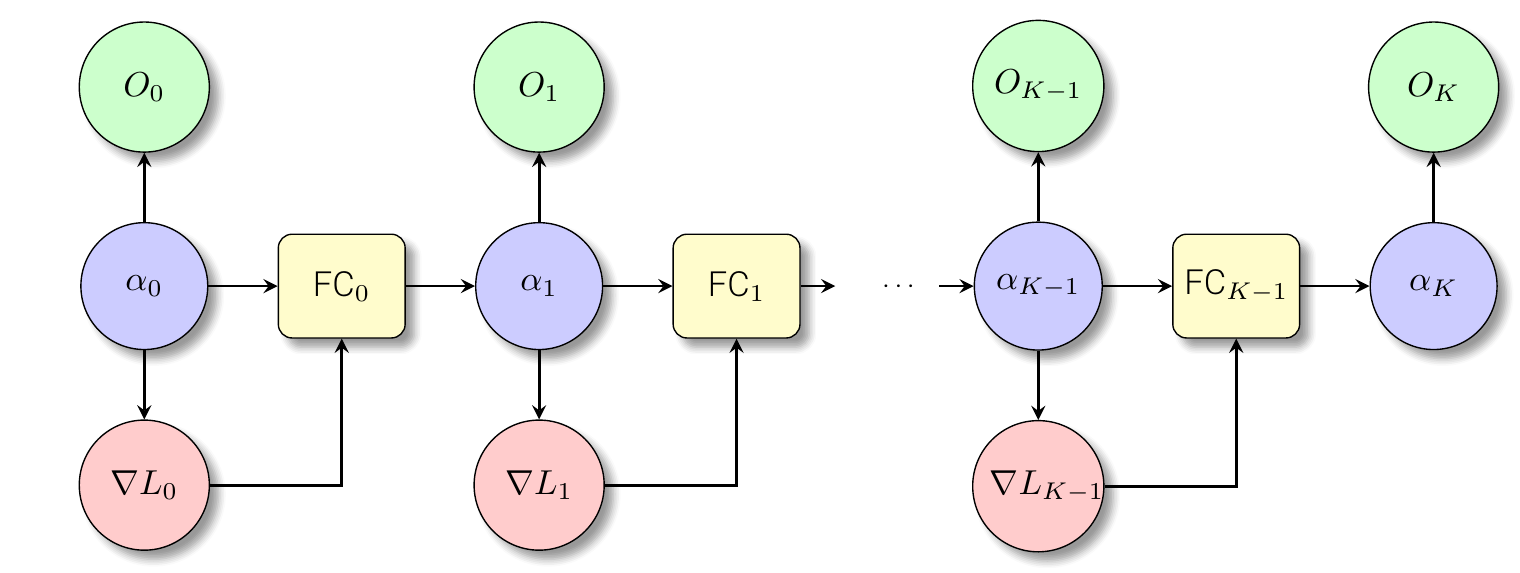}
	\caption{Scheme of the unrolled model. The estimated wavefront coefficients $\alphabold_i$ are used to compute
 the gradient of the loss function, $\nabla L_i$, which also needs the estimated object $O_i$.
 Both the coefficients and the gradients are refined by a neural network that produces updated
 wavefront coefficients. The training proceeds by forcing these neural networks to
 produce wavefront coefficients that are as close to the solution as possible at
 each step of the unrolled algorithm.}
	\label{fig:nn_unroll}
\end{figure*}

This iterative scheme is known to converge slowly. Following the inspiration
of \cite{fasta10}, we propose to modify Equation (\ref{eq:gradient_descent}) to:
\begin{equation}
	\alphabold_{i+1} = \alphabold_{i} - \eta_i F_i(\alphabold_i, \nabla L(o,\alphabold_i)),
	\label{eq:gradient_descent_neural}
\end{equation}
where the $F_i$ are (potentially different) $K$ neural networks that take as input
the current estimation of the wavefront coefficients and the gradient of the loss function
and output the correction to be applied to the wavefront coefficients. 
We found that the order of magnitude of the coefficients and the gradients 
were not so different, so that they enter into the neural network
without any extra compensation. Anyway, we would like to explore in the 
future the option of adding a learnable weight to the gradients to see if it improves the results.
A graphical 
representation of the unrolled model is displayed in Fig. \ref{fig:nn_unroll}.
After each iteration, one can compute a loss function $L_i$, which is computed with the current
estimation of the object $O_i$ and the wavefront coefficients $\alphabold_i$. Typically, 
the weights of the neural networks and the values of the learning
rates are trained end-to-end by minimizing the sum of these losses
for the $K$ steps. 

When carrying out the training in acceleration hardware like GPUs, one
can potentially find memory limitations. To overcome these limitations, we propose here to use
a greedy training of the neural networks, which consists of updating the weights 
of each individual neural network one after the other. To this end, after the first
step of the unrolled gradient descent method and the first neural network, one has all
the ingredients to compute the loss function. At this point, one can update the weights of 
the neural network $F_1$ using the gradient of $L_1$ with respect to the weights of the neural network.
After this step, one can update the weights of the neural network $F_2$ using the gradient of $L_2$
with respect to the weights of the neural network $F_2$ and so on. This process is repeated
until the weights of all the neural networks have been updated. The memory footprint of 
this approach is much lower than the end-to-end approach and it is also
much faster to train. Although it can 
lead to a potentially slightly lower accuracy, we verified in a limited amount
of comparisons that this lower accuracy does not significantly impact the results.

In our model, each neural network $F_i$ is a 1D convolutional neural network acting
on the time direction. The current estimation of the
wavefront coefficients and the gradient of the loss function are concatenated producing
a tensor of size $B \times 2M \times J$, where $B$ is the batch size, $M$ is 
the number of coefficients and $J$ is the number of frames. This tensor is
then passed through two 1D convolutional layers with $2M$ filters of size $3$ and stride $1$
with Gaussian Error Linear Unit activation functions \citep[GELU;][]{Hendrycks2016}. 
At the end, a final convolutional layer produces again 
a tensor of size $B \times M \times J$ with the wavefront coefficients for all
the images in the burst. After a non-exhaustive search for the optimal number of unrolling steps, we stick
to $K=10$. This gives a total amount of 816.2k trainable parameters, divided
into the ten 1D CNNs.

\begin{figure*}
	\includegraphics[width=1.0\textwidth]{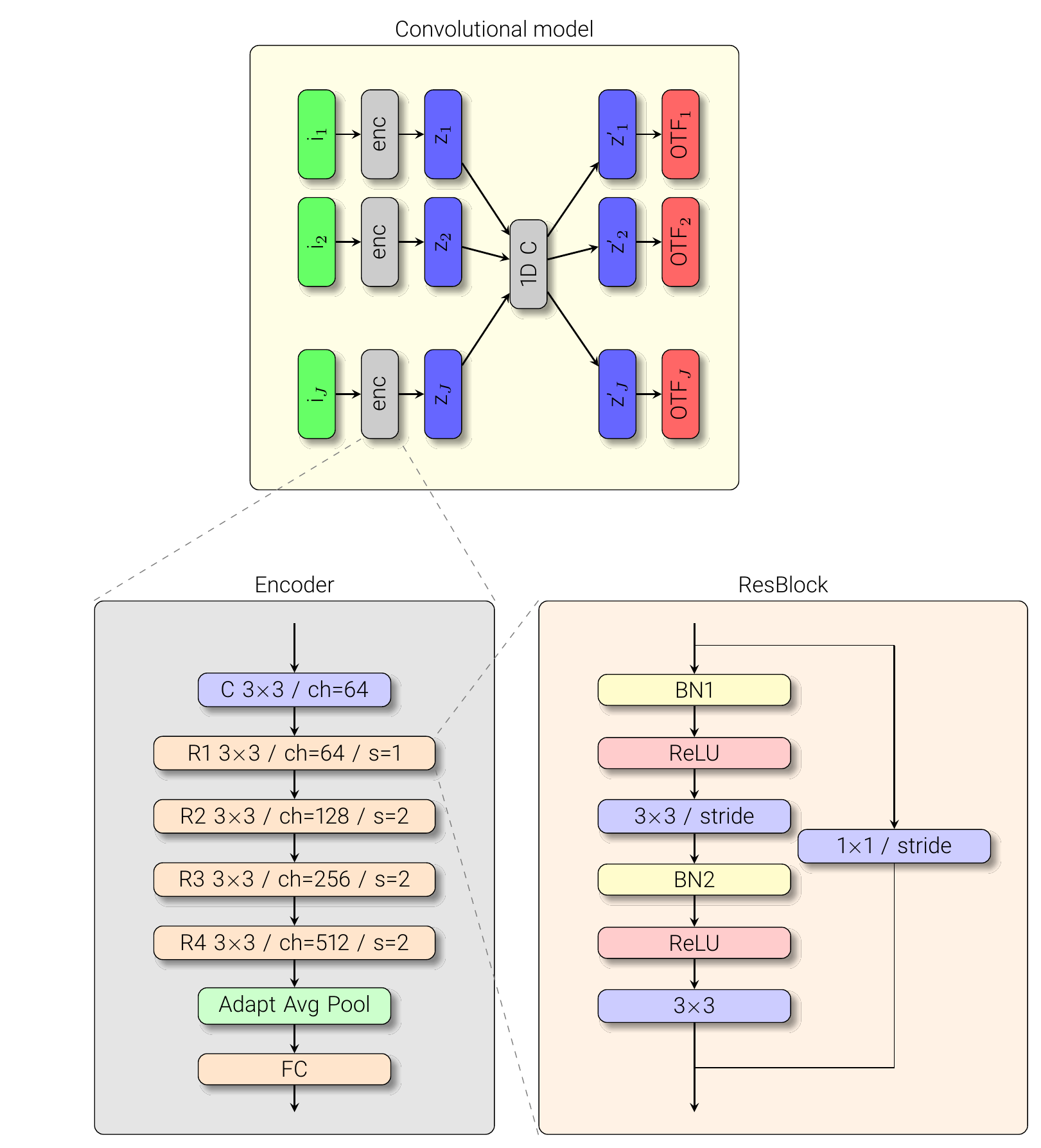}
	\caption{Convolutional model (upper panel). The Resnet18 encoder architecture is displayed
 in the gray block. The ResBlock that is part of the Resnet18 model is displayed in the
 orange block. C: convolutional layer, R: ResBlock, FC: fully connected, BN: batch normalization, 
 ReLU: rectified linear unit.}
	\label{fig:nn_cnn}
\end{figure*}

\subsection{Model 2: Fully Convolutional}
\label{sec:unsupervised}
The second model we investigate in this work is an update of the one
proposed by \cite{AsensioRamos2020}. The model is displayed in the upper
panel of Fig. \ref{fig:nn_cnn}. The observed $M$ frames are passed 
through a common convolutional encoder to extract features. We propose
to change the relatively simple encoder used by \cite{AsensioRamos2020}
to a Resnet18 \citep{DBLP:journals/corr/HeZRS15}. We found that
such a deep network is able to extract relevant features even from single
images when a second channel with a phase diversity does not exist. This neural
network is in charge of extracting relevant features that allow the model
to estimate the wavefront coefficients. Therefore, the specific solar structure
that is observed in each case has to be ignored by the neural network. 
The architecture of the Resnet18 is found in the lower left panel of Fig. \ref{fig:nn_cnn}.
An initial convolutional layer with $64$ filters of size $3 \times 3$
is followed by the consecutive application of residual blocks (ResBlock), whose
architecture is displayed in the lower right panel of Fig. \ref{fig:nn_cnn}. The
size of the input image is decreased by a factor of $2$ after each ResBlock
with a stride of 2. An adaptive average pooling produces an 
output that is independent of the size of the input images. A final
fully connected layer produces the latent vector $z_i$ for each image $i$, whose
length is equal to the number of KL modes considered in the
wavefront. We train the encoder from scratch.

Instead of combining the latent vectors $z_i$ using a Gated Recurrent Unit as proposed
by \cite{AsensioRamos2020}, we prefer to use another convolutional
neural network (attention layers are also perfectly good candidates
that need to be explored in the future). To this end, all latent vectors $z_i$ are combined into
a tensor of size $B \times M \times J$. This tensor is then passed through
a simple convolutional neural network made by the consecutive application of
two 1D convolutional layers with $2M$ filters of size $3$ and stride $1$, followed
by GELU activation functions. A final
convolutional layer produces again a tensor of size $B \times M \times J$. The role
of this CNN is to couple together all wavefront coefficients to partially take
into account their temporal correlation for all the frames in a single burst.
From these wavefront coefficients, the OTFs and the MOMFBD loss are computed.

The total number of trainable parameters is 11.2M, where roughly 99\% of the
parameters correspond to the Resnet18 encoder, while the remaining 1\%
of the parameters belong to the CNN that combines all latent vectors to
share temporal information.

\subsection{Preprocessing}
\label{set:preprocessing}
The calculation of the deconvolved image with the Wiener filter of Equation (\ref{eq:object}) requires
the Fourier transform of the observed images. Since they are not periodic, and
in order to avoid the appearance of spurious high frequencies in the final result, we
apodize them with a Hanning window of 12 pixels in each border. 

We note that it is crucial to properly normalize the input of the neural networks. In the
case of the second model, the input to the Resnet18 encoder are the 
images themselves. For this reason, we preprocess the images by transforming them 
to the $[0,1]$ interval using the minimum and maximum brightness in the burst.
The images enter into the encoder unapodized. The neural networks of 
the unrolled model deal directly with the wavefront coefficients. Specifically, these 
coefficients are never too large in the case of datasets of outstanding quality, so 
only then can the wavefront coefficients be entered into the neural networks without 
additional normalization.

In order to avoid dealing with very large tip-tilts, the observed bursts
are first destretched. This is done by computing local tip-tilts in a coarse
grid using cross-correlation. The frames are realigned by reinterpolation using bilinear
interpolation. Once the coarse destretching is complete, additional steps
with finer grids deal with the small-scale alignment. Although it is not
specially time-consuming and results almost negligible in the case of
classical deconvolution algorithms, this prealignment takes a large fraction of the computing
time when using our neural approaches. We plan to train neural networks in the future
to estimate the optical flow for destretching and to carry out this step as close
to real-time as possible.

\subsection{Tip-tilt Handling}
\label{sec:tip_tilt}
In general, and specially when the observed objects fill the entire FoV, 
it is not possible to infer an absolute tip-tilt. The fundamental reason
is that tip-tilt enters the OTF as a phase factor, thus it is obvious that 
only relative tip-tilt between different frames can be estimated. Fortunately, this can be
solved using different strategies. If sufficiently long bursts are obtained, 
one can force the mean tip-tilt to be zero and obtain the relative tip-tilts
of all frames with respect to the mean. Another option is to set the
tip-tilt of the first (or any) frame to zero and infer the tip-tilt of every
frame with respect to the first one (or the one selected as reference). We
have chosen the first option, which gives good results.

\section{Training}
\label{sec:training}
Both neural models are trained using a combination of data
coming from the SST and GREGOR. Since the physics of image formation
is part of the loss function, both models can be trained fully
unsupervisedly using only observations, following the same strategy
described by \cite{AsensioRamos2020} describe the training data in the following.

\subsection{SST Data}
\label{sec:sst}
We used high-spatial-resolution observations of diverse structures taken 
with the SST. These 
datasets were obtained in common photospheric and chromospheric diagnostics during 
our 2019 and 2020 campaigns. Table 1 outlines observing details of each dataset.

\begin{table}
\caption{Observing details of the SST datasets: date, target (QS: Quiet Sun, EN: Enhanced Network, AR: Active Region, PL: Plage), FoV center at the beginning of the observation, observed spectral lines,  temporal cadence, and for what the data were used.}
\label{tab:obs_sst_1}
\begin{tabular}{lccccc} % define the column alignment
                                  % l: left, c: center, r: right
                                  % @{.} replace the inter-column by a .
  \hline
Date & Target & Position & Spectral lines & Cadence & Use \\
 & & (x, y) & & [s] & \\
  \hline
01--Aug--19 & QS & (1", 0") & Ca \textsc{ii} 854.2 nm  & 31 & Training \\
01--Aug--19 & QS & (2", --58") & Ca \textsc{ii} 854.2 nm & 31 & Validation \\
01--Aug--19 & EN & (110", 10") & Ca \textsc{ii} 854.2 nm & 31 & Training \\
26--Jul--20 & AR & (-220", --416") & Fe \textsc{i} 617.3 nm & 60 & Training\\
 & & & Ca \textsc{ii} 854.2 nm & 60 & \\
 & & & Ca \textsc{ii} K 393.4 nm & 6 & \\
27--Jul--20 & AR & (-20", --416") & Ca \textsc{ii} 854.2 nm$^\star$ & 20 & Validation\\
 & & & Ca \textsc{ii} K 393.4 nm & 6 &\\
06--Aug--20 & PL & (-115", 315") & Fe \textsc{i} 617.3 nm & 50 & Training\\
 & & & Ca \textsc{ii} 854.2 nm & 50  & \\
 & & & Ca \textsc{ii} K 393.4 nm & 9  & \\ 
  \hline
\end{tabular}
\end{table}

We acquired spectropolarimetric data in the Fe \textsc{i} 617.3 nm and Ca \textsc{ii} 854.2 nm spectral 
lines with the CRisp Imaging Spectropolarimeter \citep[CRISP;][]{2006A&A...447.1111S, 2008ApJ...689L..69S}. 
Among other elements, its optical system includes three synchronized CCD cameras: two cameras 
collect pairs of narrowband images with orthogonal polarization states while another 
camera records wideband images that offer context information and act as anchor images 
during the restoration process with the MOMFBD technique. All cameras obtain images 
with short time exposure (17 ms) to record data where seeing-induced distortions 
are frozen. The sampling of each spectral line was 
narrow at the line core and broader at the line wings (see Table 2). We note that 
we used two different samplings for the Ca \textsc{ii} 854.2 nm line. The temporal cadences 
specified in Table 1 thus result from the number of accumulations (in Table 2), the 
exposure time, and the acquisition of four modulation states. Each CRISP image has a 
FoV of 55''$\times$55'' ($\sim$960$\times$960 pixels) with a plate scale of 0.057''.

\begin{table}
\caption{Sampling of the spectral lines observed with CRISP and CHROMIS and the number of accumulations. Wavelength positions within the square brackets are indicated from the line center.}
\label{tab:obs_sst_2}
\begin{tabular}{lcc} % define the column alignment
                                  % l: left, c: center, r: right
                                  % @{.} replace the inter-column by a .
  \hline
Spectral lines & Acc. & Wavelength positions [m\AA]\\
  \hline
Fe \textsc{i} 617.3 nm & 12 & [0, $\pm$30, $\pm$60, $\pm$90, $\pm$120, $\pm$150, $\pm$180] \\
Ca \textsc{ii} 854.2 nm & 12 & [0, $\pm$65, $\pm$130, $\pm$195, $\pm$260, $\pm$325, $\pm$390, $\pm$520, \\
 & & $\pm$650, $\pm$845, $\pm$1.040, $\pm$1.755] \\
Ca \textsc{ii} 854.2 nm$^\star$ & 12 & [0, $\pm$65, $\pm$130, $\pm$260, $\pm$520, $\pm$780, $\pm$1.040, $\pm$1.755] \\
Ca \textsc{ii} K 393.4 nm & 15 & [0, $\pm$65, $\pm$130, $\pm$195, $\pm$260, $\pm$325, $\pm$390, $\pm$455, \\
 & & $\pm$520, $\pm$585, $\pm$650, $\pm$845, $\pm$1235] + extra point at 400 nm\\
  \hline
\end{tabular}
\end{table}

The Ca \textsc{ii} K data were obtained with the CHROMospheric Imaging Spectrometer 
\citep[CHROMIS;][]{2017psio.confE..85S}. In contrast to CRISP, CHROMIS only provides 
intensity data at different wavelength positions so far. The optical system of CHROMIS also contains three CCD cameras, 
of which one collects narrowband images while the other two record wideband data. 
Specifically, one of the wideband cameras also collects data with a defocus (to
carry out deconvolution using phase diversity), which turns out to be
essential for improving the image restoration quality \citep{2021A&A...653A..68L}. 
We do not make use of this channel in this work, but the neural models can be 
trivially modified to include it. CHROMIS images have also a short time exposure 
(1--2 ms). Their FoV is of 72''$\times$45'' ($\sim$1900$\times$1200 pixels) with a pixel size of 0.038''.

We applied a basic data reduction, i.e., images in each dataset are aligned and 
corrected for dark and flat-field. In the case of the Ca \textsc{ii} 854.2 nm data, we performed 
additional processing to the flat-field data because these datasets show the imprint of
the camera electronics, showing a circuit-like pattern. This pattern is visible in infrared 
observations acquired until August 2022 
because of the decrease in the efficiency of the CCD cameras at such wavelengths. A 
traditional flat-field correction is insufficient to eliminate this pattern, so flat-field 
data processed considering the backscatter problem are necessary. More details on the 
backscatter problem of the CCD cameras can be found in \citet{2013A&A...556A.115D}.

\subsection{GREGOR Data}
\label{sec:gregor}
High-spatial-resolution H$\alpha$ filtergrams were acquired in 2022 with the improved 
High-resolution Fast Imager \citep[HiFI;][]{hifi2023}. 
The instrument consists of two synchronized CMOS cameras with $1368 \times 1040$ pixels, one narrow- and one broadband H$\alpha$ filter, which are 
attached to the GREGOR solar telescope.
This setup enables MOMFBD restoration using the information of two imaging channels. 
The spatial sampling was 0.050'' pixel$^{-1}$. The images were acquired with a frame rate of 100~Hz and an exposure time
of 9~ms. The observing strategy consisted in acquiring bursts of 500 images in about 6~s, followed by a break of 5~s, 
until the next burst started. Hence, the effective cadence is $\sim$11~s. 
The images were corrected for dark- and flat-field using sTools \citep{stools2017}, a data reduction pipeline. 
The sTools pipeline also performed image selection, retaining the best 100 images from each 500-image burst. 
The resulting 100 images were aligned and restored using MOMFBD to produce one final narrow- and one 
broadband high-resolution filtergram. The data used for training and validation is
shown in Tab. \ref{tab:obs_gregor}.

\begin{table}[!t]
\caption{Observing details of the HiFI datasets at GREGOR: date, target (F: Filament, AR: Active Region), 
FoV center at the beginning of the observation, observed wavelength, temporal cadence, and for what the data were used.}
\label{tab:obs_gregor}
\begin{tabular}{lccccc} % define the column alignment
                                  % l: left, c: center, r: right
                                  % @{.} replace the inter-column by a .
  \hline
Date & Target & Position & Wavelength & Cadence & Use \\
 & & (x, y) & & [s] & \\
  \hline
07--Jun--22 & F     & ($-594$'', $-449$'') &  H$\alpha$, H$\alpha$ continuum  & 11 & Training \\
04--Jul--22 & F, AR & ($-565$'', $235$'')  &  H$\alpha$, H$\alpha$ continuum  & 11 & Training  \\
08--Jul--22 & F, AR & ($-586$'', $201$'')  &  H$\alpha$, H$\alpha$ continuum  & 11 & Training \\
28--Nov--22 & F     & ($226$'', $-199$'')  &  H$\alpha$, H$\alpha$ continuum  & 11 & Validation \\
  \hline
\end{tabular}
\end{table}

\subsection{Training Details}
A total of 75k patches of $64 \times 64$ pixels are randomly extracted from the FoVs
of CRISP, CHROMIS, and HiFI. For CRISP and CHROMIS, wavelength position is also chosen randomly. 
For the polarization observations with CRISP, the camera and modulation states are also chosen randomly.
A dedicated training with a much larger training set should be carried out 
when these neural models are deployed for routine on-site correction of solar images.

\begin{figure}
\centering
	\includegraphics[width=0.8\textwidth, clip]{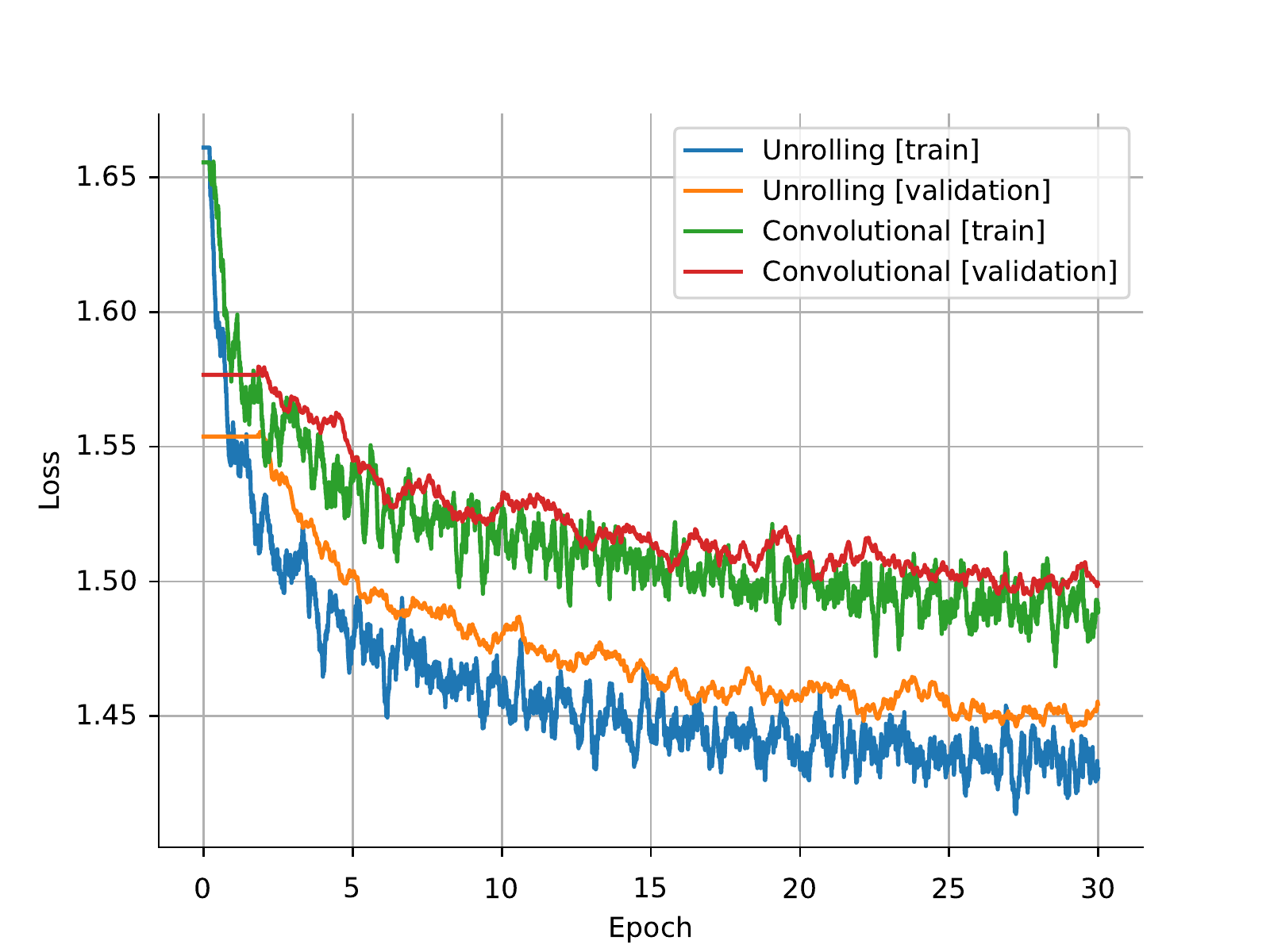}\\
	\caption{Training and validation losses for the two models considered in this 
 work. We see no hints of overfitting.}
	\label{fig:losses}
\end{figure}

The OTFs are computed from the autocorrelation of the generalized
pupil function with a wavefront calculated with the KL expansion with 44 modes. 
The computation of the OTFs and the convolution of the estimated
object and the instantaneous PSFs are computed using the Fast Fourier Transform (FFT).
The training time for one epoch was $\sim$40 min for the case of the
unrolling model and $\sim$28 min for the fully convolutional model. 
Note that, despite having more than 10 times more free parameters, the convolutional
model is faster to evaluate. This is a consequence of the fact that computing the
gradient of the loss function with respect to the wavefront coefficients for 
the unrolling model requires the computation of additional Fourier transforms.

We trained both models for 30 epochs. The training and validation losses are shown in Fig. \ref{fig:losses}
for both models. The validation loss saturates close to the end of the training, showing
no signs of overfitting.
Consequently, we choose the model parameters at the end of the training. 
The results show a slightly better training and validation loss for the unrolled
model. We did not carry out any hyperparameter tuning because of computational reasons.
We argue that the losses of both models can be made more similar and reach
lower values after a suitable hyperparameter tuning phase.

The training of our model was performed on a single NVIDIA 
GeForce RTX 2080 Ti GPU. The neural network, as well as the computation of the loss
function, was implemented and run using PyTorch 1.12 \citep{pytorch19}. 
We used the Adam optimizer \citep{Kingma2014} with a learning rate of $3\cdot 10^{-4}$. 

At test time, the unoptimized correction of a full 1k$\times$1k image carried out in
overlapping patches of 64$\times$64 takes only $\sim$8 seconds in a single 2080 Ti GPU for the
unrolled model and $\sim$6 seconds for the convolutional model.
This includes all I/O operations between the CPU and the GPU. In order to minimize the
effect of the apodization while still being computationally efficient, the patches
overlap 38 pixels, which is an area slightly larger than the chosen apodization window.
The final image is constructed by mosaicking, built 
by averaging pixels from surrounding
patches not taking into account the apodized regions. On average, our models can do the 
reconstruction of 64$\times$64 patches in $\sim$10 ms. 

\begin{figure*}
\centering
\includegraphics[width=0.65\textwidth, clip]{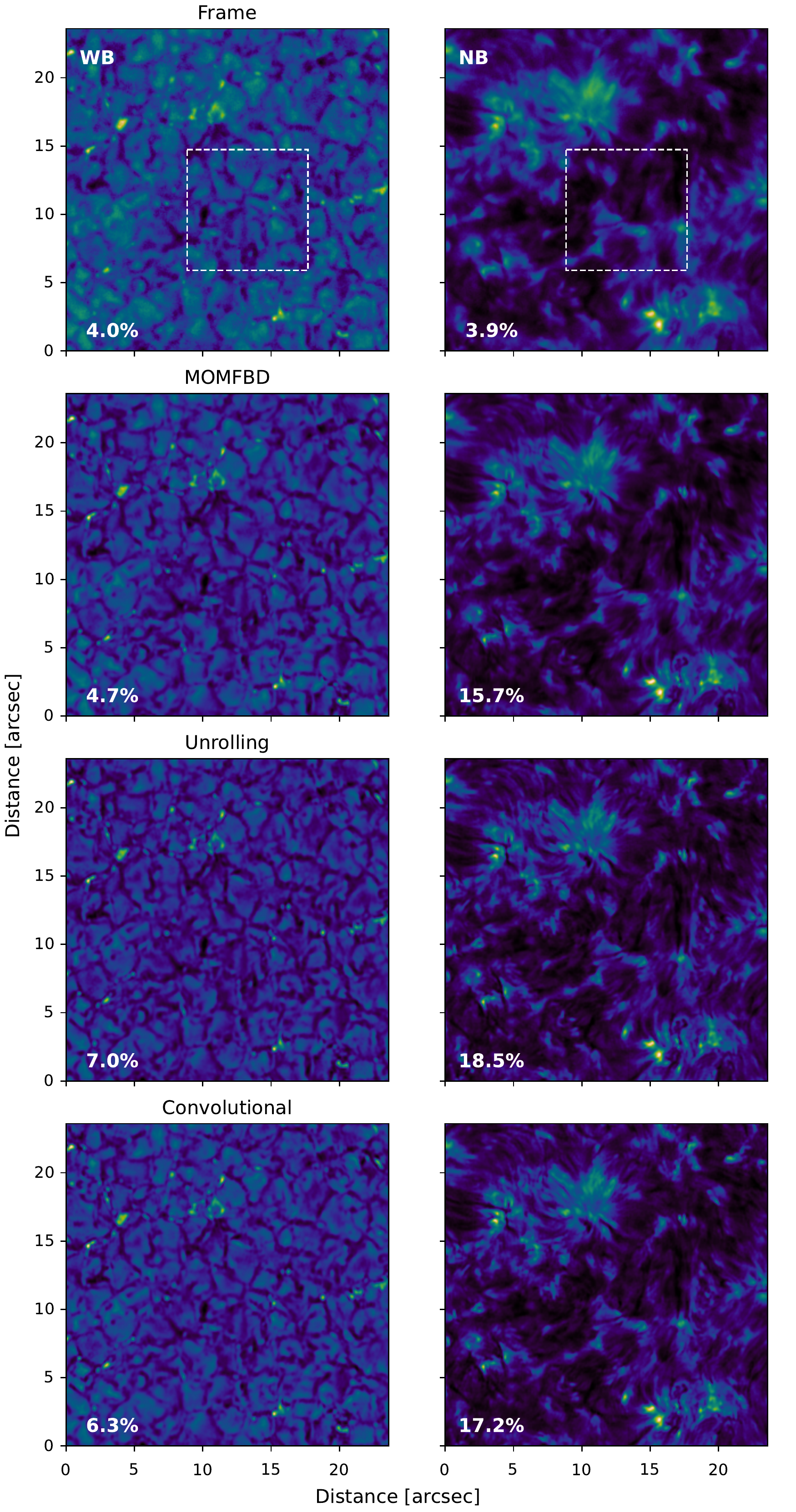}\\
\caption{Results of the restoration of an observation of the quiet Sun with CRISP
in the Ca \textsc{ii} 8542 \AA\ line taken on 01-Aug-19. The left column
shows wideband images, while the right column displays the narrowband images corresponding to 
a wavelength displacement of +65 m\AA\ with respect to line center, which
clearly shows conspicuous chromospheric structures. The contrast of the image shown
in the image is computed on the displayed rectangle.}
\label{fig:qs_8542}
\end{figure*}

\section{Restoration}
\label{sec:results}

\begin{figure*}
\centering
\includegraphics[width=0.65\textwidth, clip]{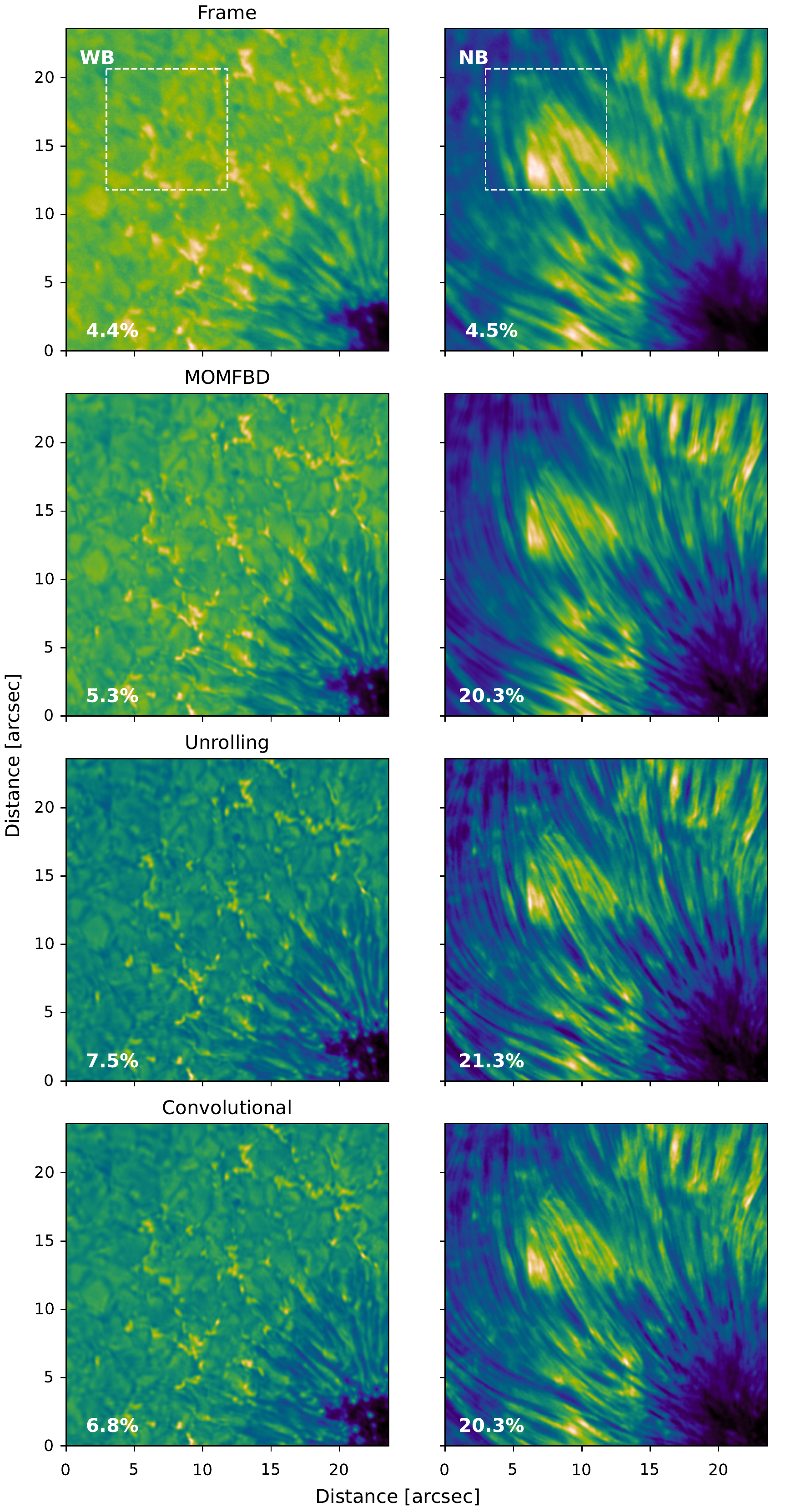}\\
\caption{Same as Fig. \ref{fig:qs_8542} but for the active region observed on 27-Jul-20.
The narrowband image is chosen at +130 m\AA.}
\label{fig:spot_8542}
\end{figure*}

\begin{figure*}
\centering
\includegraphics[width=0.65\textwidth, clip]{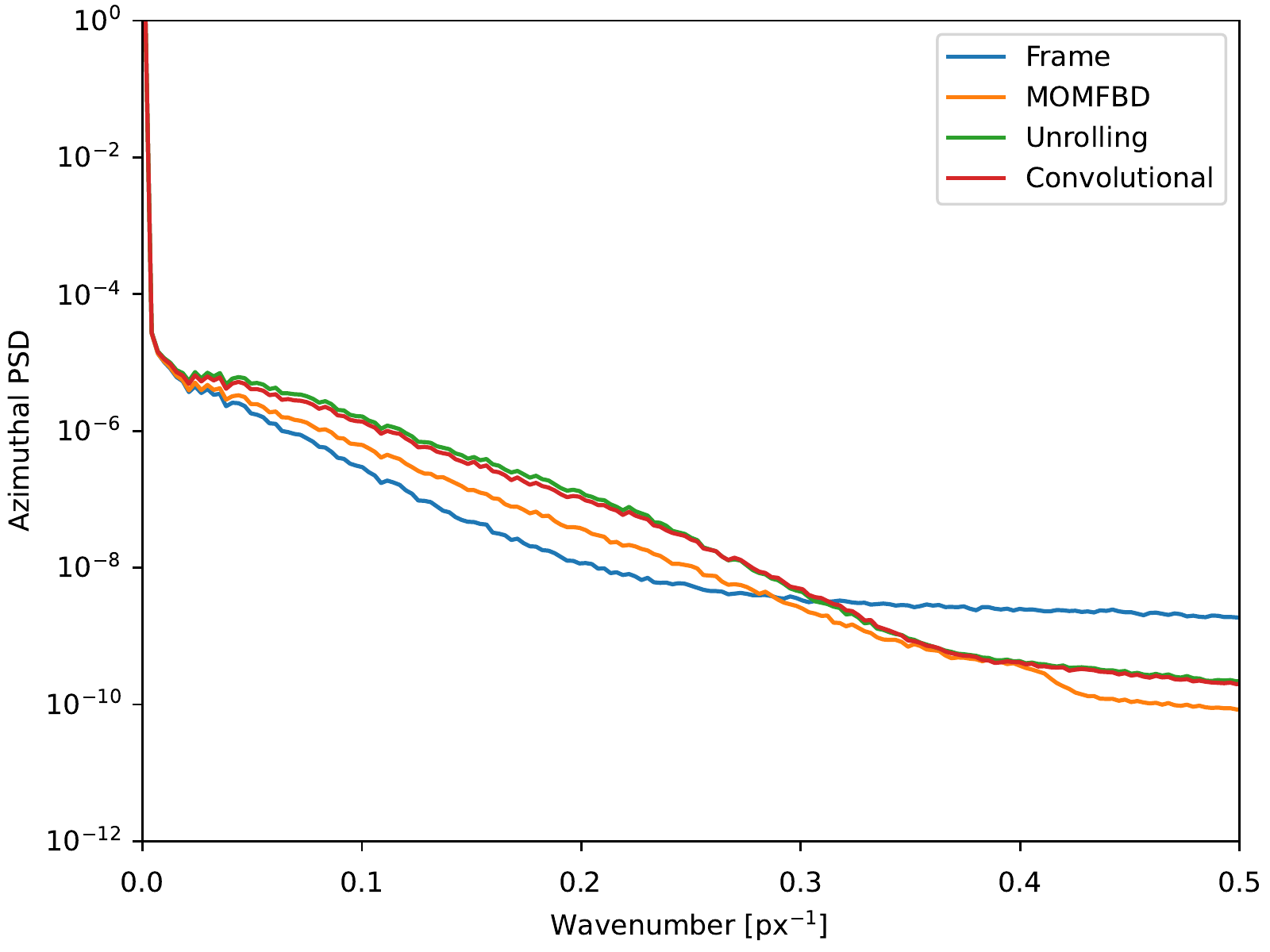}\\
\caption{Spatial azimuthal averaged power spectrum of the quiet Sun data shown in Fig. \ref{fig:qs_8542} for
the wideband data.}
\label{fig:qs_power_8542}
\end{figure*}

\subsection{CRISP}
Once the models are trained, we validate our results by applying them to observations that 
were not part of the training set. We compare the results with the application of
the classical MOMFBD code, which is performed in patches of 92$\times$92. 
Figures \ref{fig:qs_8542} and \ref{fig:spot_8542} show 
the results for CRISP observations in a quiet region and in an active region, respectively.
Both observations were carried out in the Ca \textsc{ii} 8542 \AA\ line.
The figure shows the wideband images in the left column and the narrowband images
in the right column. The top row shows the first frame of the burst, which gives
an idea of the quality of the observations. The second row displays the results
obtained with the classical MOMFBD code, while the remaining two rows show the
results of our two models. In general, the restoring capabilities of our models
are remarkable. A deeper analysis of some details individually, specially those
appearing in the narrowband channel, suggests that our results produce a crispier
image. It is well-known that the appearance of spatially correlated noise, inherent
to image deconvolution, can lead to apparently crispier images. However, we do not
see significant artifacts. It is true that the noise filter of the MOMFBD
code has been refined with the passing years and we are simply using the
recipe of \cite{lofdahl_scharmer94}. Furthermore, we see no significant visual difference between
the unrolled and the fully convolutional model.
In order to make this comparison more quantitative, we have computed the contrast (rms intensity
normalized to the average intensity) of the 
image in selected rectangles. We see that our neural approaches produce slightly larger contrasts than those
of MOMFBD.

% \begin{table}
% 	\caption{.}
% 	\label{tab:results}
% 	\begin{tabular}{lccc} % define the column alignment
% 									  % l: left, c: center, r: right
% 									  % @{.} replace the inter-column by a .
% 	  \hline
% 	Method & Instrument & Contrast & Time \\
% 	 & & \% & [s] \\
% 	\hline	
% 	Unrolling & CRISP & & \\
% 	Convolutional & CRISP & & \\
% 	MOMFBD & CRISP & & \\
% 	  \hline
% 	\end{tabular}
% 	\end{table}

Similar conclusions can be extracted from the restoration of the active region
observations displayed in Fig. \ref{fig:spot_8542} for the Ca \textsc{ii} 8542 \AA\
line. Both the unrolled and the fully convolutional models produce slightly more
crispier images than MOMFBD, where the superpenumbral filaments
in the narrowband channel sampling the chromosphere show an improved contrast. The wideband 
images also display slightly more compact structures, with the dark penumbral
filaments showing also more contrast.

\begin{figure*}
\centering
\includegraphics[width=0.65\textwidth, clip]{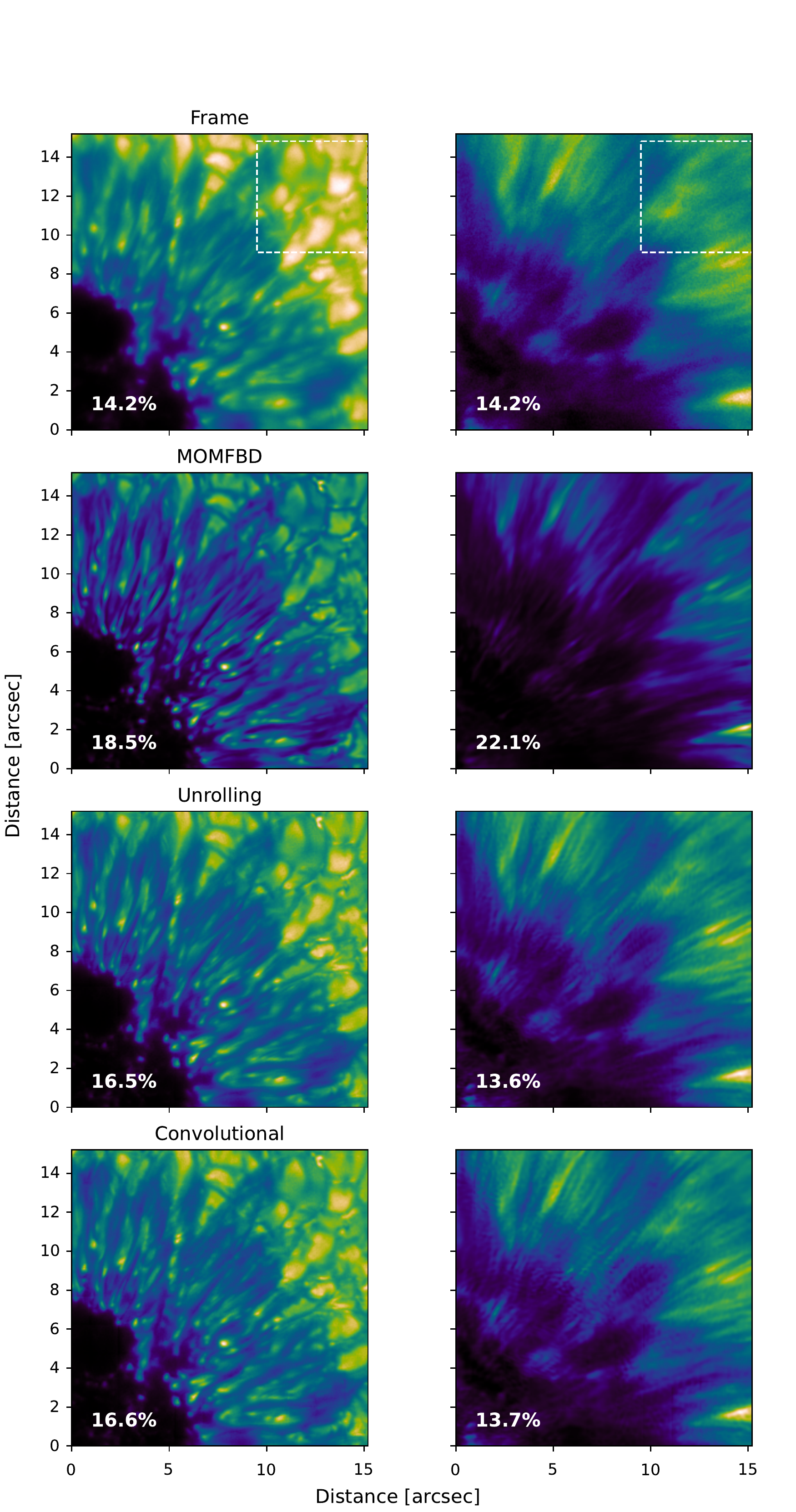}\\
\caption{Same as Fig. \ref{fig:qs_8542} but for the active region observed 
 on 27-Jul-20 with the CHROMIS instrument in the Ca \textsc{ii} K line. The
 monochromatic wavelength is also placed at +65 m\AA\ from the line center.}
\label{fig:spot_3934}
\end{figure*}

The visual impression suggested in the previous paragraphs is made more quantitative by
inspecting the azimuthal average spatial power spectrum, as displayed in Fig. \ref{fig:qs_power_8542}.
The original frames contain a significant amount of noise in the large wavenumbers
that is suitably filtered by the deconvolution process. A larger enhancement of the
power appears in the range between wavenumbers 0.02 px$^{-1}$ and 0.3 px$^{-1}$. This 
is a consequence of the deconvolution recovering structures between 0.2'' and 3'', which
are destroyed in individual frames by the instantaneous PSFs. Our two neural models are
almost equivalent and are able to extract a little more power in this range. This is the
reason why the neural reconstructions looked crispier. On the contrary, the noise
cancellation is similar between MOMFBD and our neural approaches. However, the strong
reduction of power at large wavenumbers from MOMFBD suggests that their noise filter
is performing slightly better than ours. This is not a fundamental limitation of
the neural models and both models would perform similarly to MOMFBD in terms
of noise cancellation if they eventually are part of the pipeline \citep{2021A&A...653A..68L}.

\subsection{CHROMIS}
The results of applying the neural models to CHROMIS observations close to the core of the 
Ca \textsc{ii} K line are displayed in Fig. \ref{fig:spot_3934}. The comparison of 
our results with those obtained with MOMFBD (using patches of size 88$\times$88) is 
rather unfair. The MOMFBD results 
are obtained taking into account the phase diversity channel, while ours does not
use that channel. It is well-known \cite[e.g.,][]{lofdahl_scharmer94,vannoort05}
that the addition of the defocused channel strongly constrains the results and
produces much improved results, with the disadvantage of having to deal with more 
data and having a good alignment of the two channels.
Even in this unfavorable case, our neural reconstructions produce excellent
results although the inferred contrasts are smaller than those
found with MOMFBD. Both neural approaches give comparable results. Of relevance, all structures
in the umbra are recovered with contrast very similar to that of MOMFBD.

\begin{figure*}
\centering
\includegraphics[width=0.65\textwidth, clip]{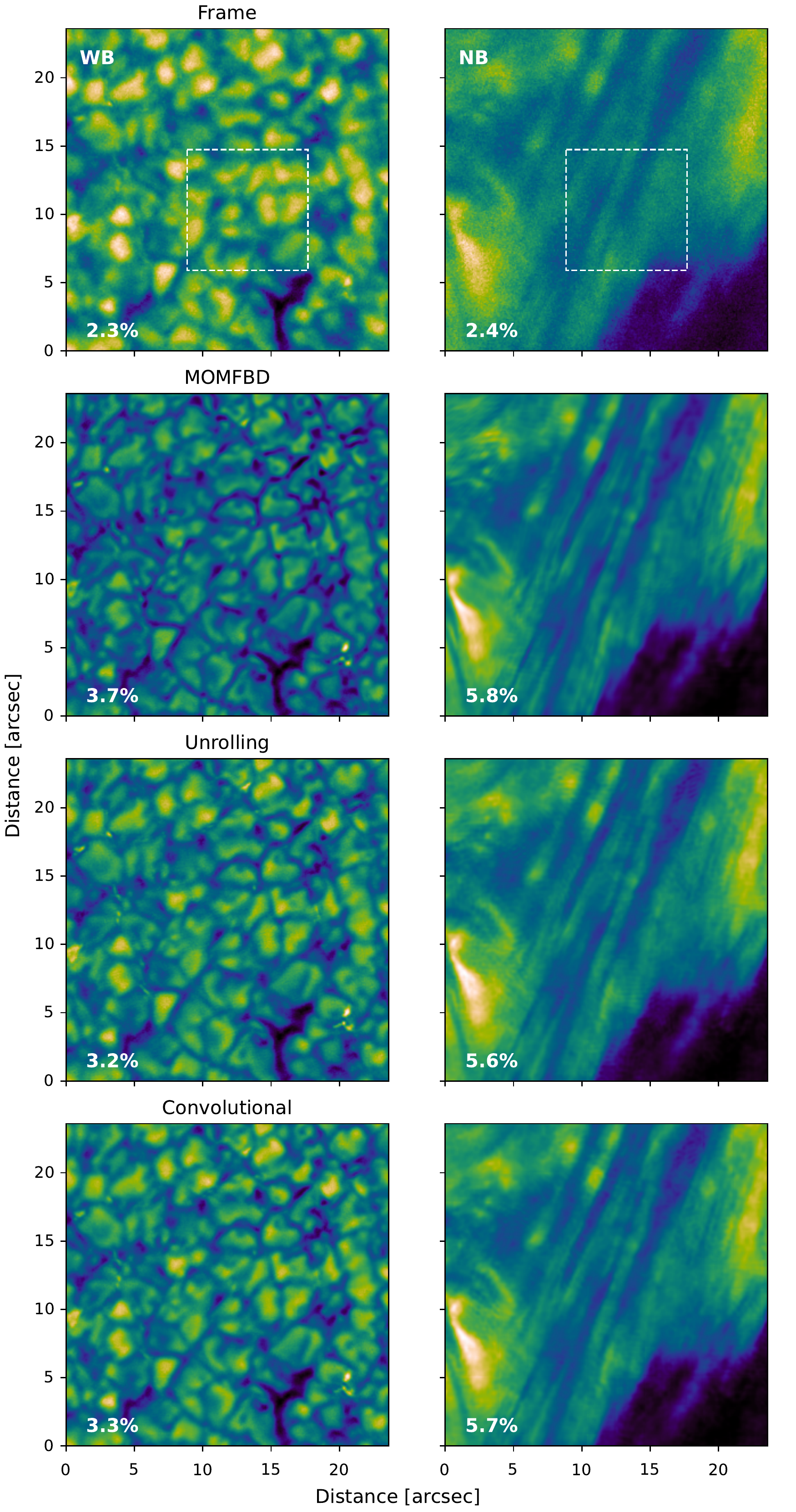}\\
\caption{Same as Fig. \ref{fig:qs_8542} but for HiFI for the observations on 28-Nov-22.}
\label{fig:qs_hifi}
\end{figure*}

\begin{figure*}
\centering
\includegraphics[width=1.\textwidth, clip]{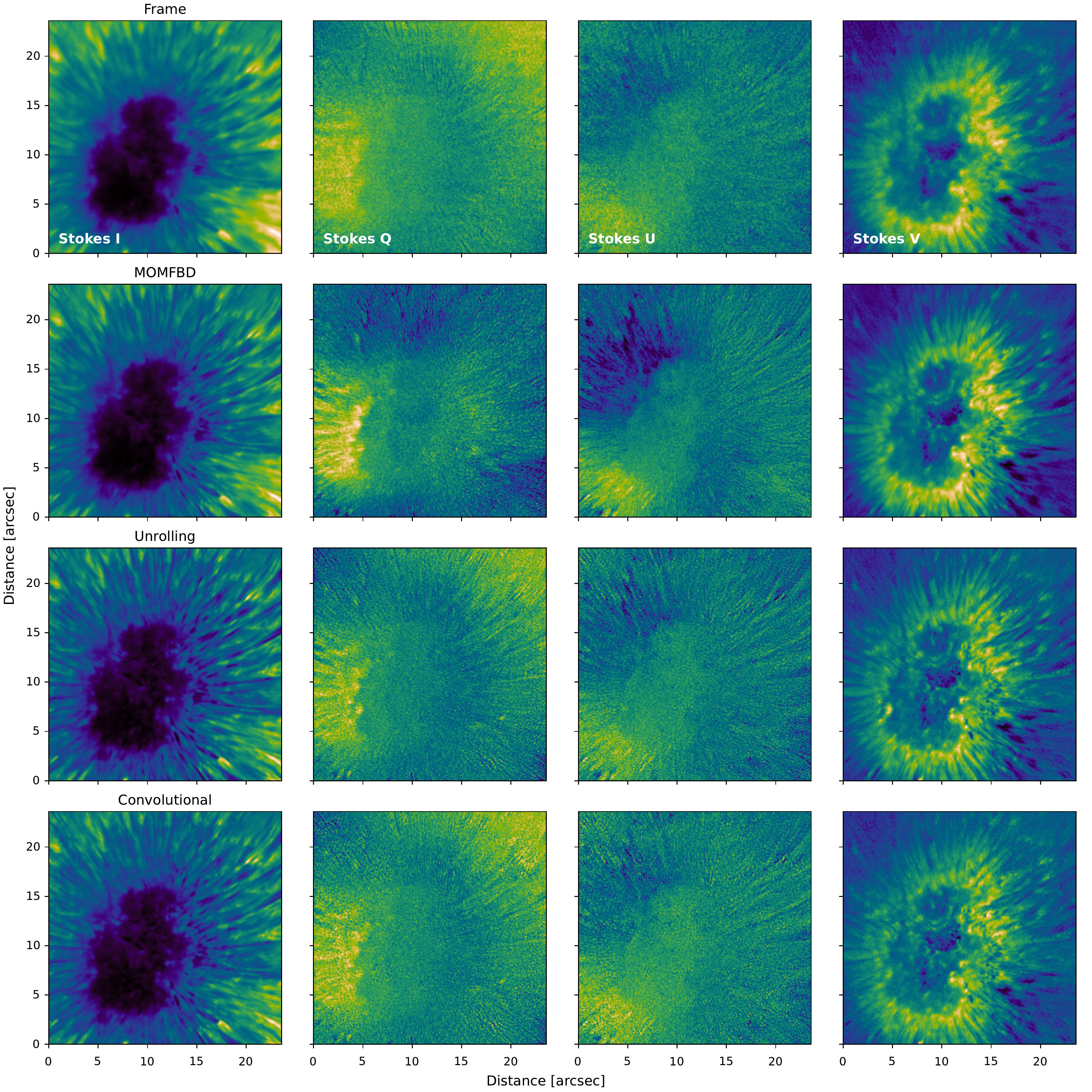}\\
\caption{Comparison of the Stokes parameter maps obtained after the polarimetric
demodulation of the original CRISP frames, together with MOMFBD and our two neural models.}
\label{fig:polarimetry}
\end{figure*}

\subsection{HiFI}
As a final application of the neural reconstruction, we show in Fig. \ref{fig:qs_hifi} a portion
of a quiet-Sun region and their reconstructions observed with the HiFI instrument compared
with that obtained with MOMFBD. The MOMFBD reconstruction was performed using 120 modes with
100 observed frames (note
that this value is higher than the number of modes we use in our neural models),
which seems to produce slightly over-reconstructed images, mainly visible
on the wideband images. The neural reconstruction
produces better-defined structures in these images, with filamentary features like
those on locations $(x,y)=(1'',17'')$ or $(4'',12'')$ being much better defined 
in the neural restorations. This might be a consequence of the strong noise filter
imposed in the MOMFBD reconstructions. The inferred contrasts are very similar
in all cases, despite using a different number of modes in the MOMFBD reconstruction.
Concerning the narrowband images, structures appear with similar quality in both
neural reconstructions. One might argue that the MOMFBD reconstructions produce slightly
more defined filaments in the upper right corner of the image. A few 
experiments carried out with this dataset shows that improvements in
the image quality are marginal after 10-20 frames, and only noise is reduced.

\subsection{Polarimetry}
Although image reconstruction produces crispier images where the effect of the
atmosphere has been reduced, it might affect the polarimetric properties of the
observations. Following the strategy of \citet{2008A&A...489..429V}, polarimetric information is recovered by modulating the measured
intensity with the aid of a polarimeter. These measurements are then demodulated
by inverting the modulation matrix. As a consequence, the Stokes parameters
are recovered as linear combinations of the measured intensities.
It is important to remember that each one of the measured modulation
states is affected by slightly different atmospheric perturbations and they are
all reconstructed independently, both in the classical MOMFBD code and in our
neural approach. As a consequence, the addition and subtraction operations
applied during the demodulation can produce artificial signals.

We use the calibrated modulation matrix for CRISP@SST to produce Stokes $I$ and 
Stokes $V$ monochromatic maps from the narrowband camera, which are displayed in Fig. \ref{fig:polarimetry}.
We see again images with a slightly better contrast with the neural approaches when
compared with MOMFBD in Stokes $I$, similar to the results of Fig. \ref{fig:spot_8542}. Concerning
the polarimetric information, let us first look at Stokes $V$, which presents
larger signals. The neural reconstruction produces more compact structures, especially
on the unrolling model. This is clearly visible in the patch at position $(5'',7'')$, which
corresponds to a penumbral grain clearly seen in Stokes $I$. The same happens
for all the opposite polarity features (probably produced by strong Doppler motions) all around 
the penumbra, in consistency with what we observe in Stokes $I$.
Concerning linear polarization, our neural models produce signals that are fundamentally 
the same signals as MOMFBD but weaker, more in accordance with the amplitudes seen in the 
original frames. Whether the MOMFBD data is affected in this case by 
deconvolution-induced crosstalk is something to be investigated in the future. The 
solution to this problem is difficult given the absence of a ground-truth solution. It is 
true, though, that our results display a little more spatially correlated noise. We anticipate
that this can be alleviated with a slight improvement on the filter used in the
Wiener estimation of the object. In general, the unrolling model produces slightly better
results.

\section{Conclusions}
\label{sec:conclusions}
We have presented two different neural models for the very fast restoration of
solar images. The first one is based on algorithm unrolling, while the second
one is an architectural improvement over the one published by \cite{AsensioRamos2020}. Both
models are trained unsupervisedly with observations from the CRISP@SST, CHROMIS@SST and
HiFI@GREGOR. Although the number of free parameters is more than an order of
magnitude higher in the convolutional model, the gradient of the loss function needs 
to be computed in the unrolled model, so the evaluation time is similar in both models. 
In any case, they are several orders of magnitude faster
in terms of computing time per anisoplanatic patch when compared with the
classic MOMFBD code. Therefore, both approaches can be seen as strong candidates for 
a near real-time image restoration method for present and future observations.

Despite the promising results shown in this paper, there are still some points 
that need to be dealt with to 
fully exploit the capabilities of machine learning in solar image restoration.
The first one is adding a phase diversity channel. This channel takes strictly
simultaneous images but a known aberration (commonly a defocus) is added. This addition
comes with the undesired effect of increasing the size of the observations, although
improved reconstructions can be obtained \citep{lofdahl_scharmer94,vannoort05}. Our training
data does not contain this phase diversity channel, but both models can seamlessly 
use it. The images of this channel are assumed to have the same aberrations as those
of the science camera, but with a known term added to Equation (\ref{eq:wavefront}). 
The unrolled model can use a defocused channel by simply taking it into account
during the computation of the gradient of the loss function. The convolutional model
can deal with the phase diversity channel by stacking this image to the input of the
Resnet18 encoder and computing the two OTFs (focused and defocused). 

The second point is that, due to a lack of computing capabilities, 
we did a non-exhaustive search for architectures and training hyperparameters. 
An exhaustive search, with the aim of producing the
most optimal architecture that leads to the best results, is left as work for 
the future.

The final point is how we can leverage machine learning to produce advances not
only in terms of computing speed. In this sense, there are still a few ideas that have never
been implemented and would improve image restoration. The first idea is
adding time evolution during the reconstruction. Currently, every burst observed at
a single scan position is deconvolved in isolation. As a consequence, once 
all bursts are deconvolved, the time coherence is a byproduct. This happens often
for observations under good seeing conditions, but it is not the case when the 
seeing conditions are worse. Deconvolving all scan positions at once is a possibility,
because one can introduce regularity constraints on the deconvolved images. A certain type
of smoothness is foreseen in the reconstructed images because one does not expect sudden 
changes from one image to the next. This is simply a consequence of the limited
sound speed in the solar atmosphere, which limits the amount of variability in
the solar images obtained with a cadence of tens of seconds.
The second idea is how to deal with spatially variant PSFs, a consequence of the 
small anisoplanatic patches in strong turbulence and/or large telescopes.
The current overlap-add (OPA) approach or mosaicking (divide the image into overlapping small
patches, which are then reconstructed and stitched back) is routinely used
in solar physics \cite[see, e.g., the excellent results of][]{vannoort05}. 
However, more precise approaches like the widespread method of \cite{nagy_oleary98} and 
the recent space-variant OLA \citep{hirsch10} methods can be used \cite[see][for a review]{denis15}.
We are currently working on a deep learning approach to the spatially variant deconvolution
that can simplify the deconvolution of whole images. When coupled together
with the models described in this paper, we will be able to propose spatially 
variant deconvolution methods that can restore whole images in a single step.

%%%%%%%%%%%%%%%%%%%%%%%%%%%%%%%%%%%%%%%%%%%%%%%%%%%%%%%%%%%%%%%%%%%%%%%%%%%
\begin{acks}[Acknowledgments]
The authors thank Michiel van Noort for invaluable advises on solar image restoration and
Nigul Olspert for his work on the initial phases of this work.
%N. Olspert thanks Michiel van Noort for supervising his postdoc at Max Planck Institute for Solar System Research.
The authors acknowledge Sergio J. Gonz\'alez Manrique, who participated in the GREGOR 
observing campaigns. The authors are also thankful to Andrea Diercke, the PI of the 
HiFI validation data, for allowing the use of the data. We also acknowledge Javier Trujillo Bueno for participating in the acquisition of the quiet-Sun data during our 2019 campaign at the SST. The Swedish 1-m Solar Telescope is operated on the island of La Palma by the Institute for Solar Physics of Stockholm University in the Spanish Observatorio del Roque de los Muchachos of the Instituto de Astrofísica de Canarias. The Institute for Solar Physics is supported by a grant for research infrastructures of national importance from the Swedish Research Council (registration number 2017-00625). The authors thankfully acknowledge the technical expertise and assistance provided by the Spanish Supercomputing Network (Red Espa\~nola de Supercomputaci\'on), as well as the computer resources used: the La Palma Supercomputer, located at the Instituto de Astrofísica de Canarias. The 1.5-meter GREGOR solar telescope was built by a German consortium under the leadership of the Leibniz Institute for Solar Physics (KIS) in Freiburg with the Leibniz Institute for Astrophysics Potsdam (AIP), the Institute for Astrophysics G\"ottingen, and the Max Planck Institute for Solar System Research (MPS) in G\"ottingen as partners, and with contributions by the Instituto de Astrof\'{\i}sica de Canarias (IAC) and the Astronomical Institute of the Academy of Sciences of the Czech Republic (ASU).
\end{acks}

%\acknowledgment US spelling: \verb+\acknowledgment+
%\acknowledgement British  spelling: \verb+\acknowledgement+

%%%%%%%%%%%%%%%%%%%%%%%%%%%%%%%%%%%%%%%%%%%%%%%%%%%%%%%%%%%%%%%%%%%%%%%%%%%

\begin{authorcontribution}
AAR proposed the two neural models, curated the training data, trained the models and analyzed the
results. AAR made all figures and wrote the text.
SEP observed and reduced the CRISP@SST and CHROMIS@SST data, also producing the MOMFBD restorations. CK
observed and reduced the HiFI@GREGOR data, also producing the MOMFBD restorations. NO did some initial 
experiments on the convolutional model. SEP, CK and NO also contributed to the manuscript.
\end{authorcontribution}

\begin{fundinginformation}
AAR acknowledges financial support from 
the Spanish Ministerio de Ciencia, Innovaci\'on y Universidades through project PGC2018-102108-B-I00 and FEDER funds. CK acknowledges funding from the European Union's Horizon 2020 research and innovation programme under the Marie Sk\l{}odowska-Curie grant agreement No 895955. SEP acknowledges the funding received from the European Research Council (ERC) under the European Union's Horizon 2020 research and innovation program (ERC Advanced grant agreement No. 742265).
\end{fundinginformation}

\begin{codeavailability}
The training and evaluation code is freely available in the following repository: \texttt{https://github.com/aasensio/neural-MFBD}.
This repository also contains information on how to retrieve the training data.
\end{codeavailability}

\begin{ethics}
\begin{conflict}
The authors declare that they have no conflicts of interest.
\end{conflict}
\end{ethics}

     % format of references provided by the journal (.bst)
\bibliographystyle{spr-mp-sola}
     % name your Bibtex file containing your references (.bib)
% \bibliography{paper}	

%	 \bibliography{paper}

     % Checking: look if the file containing the ``\bibitem'' exits
     %           so check if the .bbl file exist (bibTeX compilation)
\IfFileExists{\jobname.bbl}{} {\typeout{}
\typeout{****************************************************}
\typeout{****************************************************}
\typeout{** Please run "bibtex \jobname" to obtain} \typeout{**
the bibliography and then re-run LaTeX} \typeout{** twice to fix
the references !}
\typeout{****************************************************}
\typeout{****************************************************}
\typeout{}}

\end{article} 

\end{document}